\documentclass[aps,prd,floatfix,nofootinbib]{revtex4} 
\usepackage{graphics}
\usepackage{graphicx}

\usepackage{latexsym}
\usepackage[cp866]{inputenc}
\usepackage[english]{babel}
\input epsf

\begin{document}

\title{\boldmath Tentative estimates of
$\mathcal{B}(X(3872)\to\pi^0\pi^0\chi_{c1})$ and
$\mathcal{B}(X(3872)\to\pi^+\pi^-\chi_{c1})$}
\author{N. N. Achasov\,\footnote{achasov@math.nsc.ru}
and G. N. Shestakov\,\footnote{shestako@math.nsc.ru}}
\affiliation{\vspace{0.2cm} Laboratory of Theoretical Physics, S. L.
Sobolev Institute for Mathematics, 630090, Novosibirsk, Russia}


\begin{abstract}

The rates of the $X(3872)\to\pi^0\pi^0\chi_{c1}$ and $X(3872)\to
\pi^+\pi^-\chi_{c1}$ decays are estimated in the model of the
triangle loop diagrams with charmed $D^*\bar DD$ and $\bar D^*D\bar
D$ mesons in the loops. There are the triangle logarithmic
singularities in the physical region of the $X(3872)\to\pi^0\pi^0
\chi_{c1}$ decay which manifest themselves as narrow peaks  in the
$\pi^0\chi_{c1}$ mass spectrum near the $D^0\bar D^0$ threshold. The
model predicts approximately the same branching fractions of the
$X(3872)\to\pi^0\pi^0 \chi_{c1}$ and $X(3872)\to\pi^+\pi^-\chi_{c1}$
decays at the level of about (0.8--1.7)$\times10^{-4}$. A distinct
prediction of the model is the value of the ratio $\mathcal{R}=
\mathcal{B}(X(3872)\to\pi^+\pi^-\chi_{c1})/\mathcal{B}(X(3872)
\to\pi^0\pi^0\chi_{c1})\approx1.1$. It weakly depends on the
$X(3872)$ resonance parameters and indicates a significant violation
of the isotopic symmetry according to which one would expect
$\mathcal{R}=2$.

\end{abstract}

\maketitle

\section{Introduction}

The modern studies of the first candidate for exotic charmoniumlike
states $X(3872)$ or $\chi_{c1}(3872)$ \cite{PDG23} advance in the
line increasing the data accuracy and expanding the nomenclature of
it production and decay channels \cite{PDG23,Ab19,Bh19,Abl19,Abl20,
Aai20,Yi21,Hir23,Aa23,Tan23,Abl23,Bh16,Ab23}. For example, the
BESIII \cite{Ab19} and Belle \cite{Bh19} collaborations obtained
information about the rate for the isospin-violating decay $X(3872)
\to\pi^0 \chi_{c1}$. Also the Belle \cite{Bh16} collaboration and
recently the BESIII \cite{Ab23} collaboration obtained upper limits
on the probability of the $X(3872)\to\pi^+\pi^-\chi_{c1}$ decay
which formally preserves $G$-parity.

According to the Belle collaboration \cite{Bh16} and the Particle
Data Group \cite{PDG23}
$\mathcal{B}(X(3872)\to\pi^+\pi^-\chi_{c1})<7\times 10^{-3}$ at the
90\% confidence level (CL). According to the BESIII data
\cite{Ab19,Ab23}
\begin{eqnarray}\label{Eq1}
\mathcal{R}_1=\frac{\mathcal{B}[X(3872)\to\pi^+\pi^-\chi_{c1}]}{\mathcal{B}
[X(3872)\to\pi^+\pi^-J/\psi]}<0.18\ (90\%\ \mbox{CL})\ [13] \quad
\mbox{and} \quad \mathcal{R}_2=\frac{\Gamma(X(3872)\to\pi^+
\pi^-\chi_{c1})}{\Gamma(X(3872)\to\pi^0\chi_{c1})}<0.2\ [2,13].
\end{eqnarray} The BESIII result \cite{Ab23} for $\mathcal{R}_1$
is consistent with the measurement from the Belle collaboration
\cite{Bh16}. An upper limit on the ratio $\mathcal{R}_2$ turned out
to be two orders of magnitude smaller than the value of
$\frac{\Gamma(2^3P_1\to\pi^+\pi^-\chi_{c1})}{\Gamma (2^3P_1 \to
\pi^0\chi_{c1})}\approx25$ expected under a pure charmonium $2^3P_1$
assumption for the $X(3872)$ \cite{DV08}. Therefore, Ref.
\cite{Ab23} concluded that the BESIII data favor the nonconventional
charmonium nature of the $X(3872)$ state. But this is not quite
true. The point is that the large theoretical value for
$\mathcal{R}_2$ found in Ref. \cite{DV08} is entirely due to the
tiny ($\simeq 0.06\,\mbox{keV}$) decay width of $2^3P_1\to\pi^0
\chi_{c1}$, calculated in this work under the assumption of the
two-gluon production mechanism of the $\pi^0$, which is not a
consequence of the hypothesis about the nature of the $X(3872)$. The
mechanism of the isospin-violating decay of $2^3P_1\to gg\chi_{c1}
\to\pi^0 \chi_{c1}$ considered in Ref. \cite{DV08} is not a single
one, and much less the leading one, for the $2^3P_1$ charmonium
state with a mass of 3872 MeV. The now known value for the decay
width $\Gamma(X(3872)\to\pi^0\chi_{c1})=(0.04\pm0.02)\ \mbox{MeV}$
\cite{Ab19,PDG23} can be explained, for example, by the mechanism of
the $2^3P_1$ $c\bar c$ $X(3872)$ state transition into $\pi^0\chi_{
c1}$ via the intermediate $D^*\bar DD^*$ and $\bar D^*D\bar D^*$
mesonic loops, see Ref. \cite{AS24} and references herein. Thus, the
results of the BESIII collaboration \cite{Ab23} have yet to be
compared with the assumed possible variants for the nature of the
$X(3872)$ state.

In anticipation of future experiments on the decays $X(3872)\to
\pi^0\pi^0\chi_{c1}$ and $X(3872)\to\pi^+\pi^-\chi_{c1}$, it is
interesting to estimate their probabilities and, accordingly, the
deviation from the relation $\mathcal{B} (X(3872)\to\pi^0 \pi^0
\chi_{c1})=\frac{1}{2}\mathcal{B} (X(3872)\to\pi^+\pi^- \chi_{c1})$
that takes place in the unbroken isotopic symmetry. These estimates
are the subject of this work.

Earlier in the work \cite{FM08}, with the use a combination of the
heavy hadron chiral perturbation theory and effective field theory
for the X(3872), the following results were obtained:
\begin{eqnarray}\label{Eq2}
\left(\frac{\mathcal{B}[X(3872)\to\pi^0\pi^0\chi_{c1}]}{\mathcal{B}
[X(3872)\to\pi^0\chi_{c1}]}\right)_{LO}=6.1\times10^{-1}, \quad
\left(\frac{\mathcal{B}[X(3872)\to\pi^+\pi^-\chi_{c1}]}{\mathcal{B}
[X(3872)\to\pi^0\chi_{c1}]}\right)_{LO}\approx\mathcal{O}(10^{-3}).
\end{eqnarray}
These estimates were performed with accounting the contributions of
the leading order (LO) diagrams for the amplitudes of the
transitions $D^0\bar D^{*0}\to\pi^0\chi_{c1}$ and $D^0\bar
D^{*0}\to\pi\pi\chi_{c1}$ \cite{FM08}. Subsequently, the value of
the ratio $\left(\frac{\mathcal{B}[X(3872)\to\pi^0\pi^0 \chi_{c1}]}{
\mathcal{B}[X(3872)\to \pi^0 \chi_{c1}]}\right)_{LO}$ was adjusted
towards its decrease by two orders of magnitude as a result of
recalculation of the $X(3872)\to\pi^0\pi^0\chi_{c1}$ amplitude
\cite{FM12}:
\begin{eqnarray}\label{Eq3}
\left(\frac{Br[X(3872)\to\chi_{c1}\pi^0\pi^0]}{Br[X(3872)\to\chi_{c1}
\pi^0]}\right)_{LO}=2.9\times10^{-3}. \end{eqnarray}

In the present work (as in Refs. \cite{AS24,AR14,AR15,AR16,AS19,
AKS22}), we consider the $X(3872)$ meson as a $\chi_{c1}(2P)$
charmonium state which has the equal coupling constants with the
$D^{*0}\bar D^0$ and $D^{*+}D^-$ channels owing to the isotopic
symmetry. Its decay into $D^*\bar D+c.c.$ occurs [similarly to, for
example, the $\psi(3770)\to D\bar D$ decay] by picking up of a light
$q\bar q$ pair from vacuum quark-antiquark fluctuations, $c\bar
c\to(c\bar q)(q\bar c)\to D^*\bar D+c.c.$. Undoubtedly, the main
feature of the $X(3872)$ resonance is that it is located directly at
the threshold of its main decay channel into $D^{*0}\bar D^0+c.c.\to
D^0\bar D^0\pi^0$ \cite{PDG23}. This circumstance ensures the
smallness of its width (it is $\sim1$ MeV) and clear violation of
the isotopic symmetry  against a background of the kinematically
closed decay channel of the  $X(3872)$ into $D^{*+}D^- +c.c.$ (the
thresholds of the $D^{*0}\bar D^0$ and $D^{*+}D^-$ channels are
separated by 8.23 MeV). Section II considers the kinematics of the
decays $X(3872)\to\pi^0 \pi^0\chi_{c1}$ and $X(3872)\to\pi^+
\pi^-\chi_{c1}$. Section III discusses hadronic loop diagrams, which
we use to estimate the branching fractions of these processes. The
estimates themselves are given in Sec. IV. Conclusions from the
analysis performed are presented in Sec. V.


\section{\boldmath Kinematics of the $X(3872)\to\pi\pi\chi_{c1}$
decays}

Let us use the Particle Data Group data \cite{PDG23} and put a mass
of the $X(3872)$ state equal to $m_X=3871.65$ MeV, and also
$m_{\chi_{c1}}=3510.67 $ MeV, $m_{\pi^+}=139.57039$ MeV, and
$m_{\pi^0}=134.9768$ MeV. The invariant phase volumes (PV)
\cite{BK73} for the three-body decays
$X(3872)\to\pi^0\pi^0\chi_{c1}$ and $X(3872)\to\pi^+ \pi^-\chi_{c1}$
are equal to
\begin{eqnarray}\label{Eq4}
\mbox{PV}(m_X;m_{\pi^0},m_{\pi^0},m_{\chi_{c1}})=0.0049718\
\mbox{GeV}^2,\ \ \ \mbox{PV}(m_X;m_{\pi^+},m_{\pi^-},m_{\chi_{c1}})=
0.00407956\ \mbox{GeV}^2, \end{eqnarray} respectively. For
comparison, we point out that the invariant phase volumes for the
decays $X(3872)\to D^0\bar D^0\pi^0$ and $X(3872)\to\pi^+\pi^-J/
\psi$ are equal to 0.0000686751 and  0.225852 GeV$^2$, respectively.
The energy release in $X(3872)\to \pi^0\pi^0\chi_{c1}$ is $\,
\mbox{T}_n=m_X-2m_{\pi^0}-m_{\chi_{c1}}= 91.0264\ \mbox{MeV}$, and
that in $X(3872)\to \pi^+\pi^-\chi_{c1}$ $\ \mbox{T}_c =m_X-2
m_{\pi^\pm}- m_{\chi_{c1}}=81.8392\ \mbox{MeV} $. The invariant mass
of the $\pi^0\chi_{c1}$ system, $m_{\pi^0 \chi_{c1}}$, varies from
$m_{\chi_{c1}}+m_{\pi^0}$ to $m_X -m_{\pi^0 }$, i.e., in the
near-threshold region with a width of 91.0264 MeV, and the invariant
mass of the $\pi^\pm\chi_{c1}$ system, $m_{\pi^\pm \chi_{c1}}$,
varies from $m_{\chi_{c1}}+m_{\pi^\pm}$ to $m_X-m_{ \pi^\mp}$, i.e.,
in that with a width of 81.8392 MeV. It is quite natural to believe
that in these regions the production amplitudes of the
$\pi^0\chi_{c1}$ and $\pi^\pm \chi_{c1}$ pairs will be dominated by
contributions from the corresponding lower partial waves.

Let us denote the four-momenta of the particles in the decay
$X(3872)\to\pi\pi\chi_{c1}$ as $p_X=p_1$, $p_{\chi_{c1}}=p_2$,
$p_{\pi_1}=p_3$, $p_{\pi_2}=p_4 $, where $\pi_1=\pi^0_1\ \mbox{or}\
\pi^+$ and $\pi_2=\pi^0_2\ \mbox{or}\ \pi^-$, and the polarization
four-vectors of the $X(3872)$ and $\chi_{c1}$ mesons as
$\varepsilon_X=\varepsilon_1$ and $\varepsilon_{\chi_{c1}}=
\varepsilon_2$. The matrix element $\mathcal{M}$ of the decay
$X(3872)\to\pi\pi \chi_{c1}$ is described in general case by five
independent invariant amplitudes $b_{i=1, ...,5}$ and it can be
written as: \begin{eqnarray}\label{Eq5}
\mathcal{M}=\varepsilon^\mu_1\varepsilon^{\nu*}_2\mathcal{M_{\mu\nu}
}=\varepsilon^\mu_1\varepsilon^{\nu*}_2\left(g_{\mu\nu}b_1+p_{2\mu}
p_{1\nu}b_2+\Delta_\mu\Delta_\nu b_3+\Delta_\mu p_{1\nu}b_4+p_{2\mu}
\Delta_\nu b_5\right),\end{eqnarray} where $\Delta=p_3-p_4$; $b_i
=b_i(m^2_X;s,t,u)$, $s=(p_2+p_3)^2=(p_1-p_4)^2$, $t=(p_2+p_4 )^2=
(p_1-p_3)^2$, $u=(p_3+p_4)^2=(p_1-p_2)^2$, and $s+t+u=m^2_X+m^2_{
\chi_{c1}}+2m^2_\pi$. Here we indicated the dependence of the
invariant amplitudes from $m^2_X$ because in what follows we will
need to replace $m^2_X$ in $\mathcal{M}$ with the variable quantity
$S_1$ meaning the invariant mass squared of the virtual $X(3872)$
state.

The $\pi\pi$ system in the $X(3872)\to\pi\pi\chi_{c1}$ decay has the
positive $C$ parity. As a consequence, only even orbital moments are
allowed in this system and states with the isospin $I=1$ are
forbidden. It is clear that the matrix element $\mathcal{M}$ must be
an even function of $\Delta$, i.e., should not change with the
permutation of $p_3$ and $p_4$, and the invariant amplitudes must
possess the following crossing properties: $b_{1,2,3
}(m^2_X;s,t,u)=b_{1,2,3}(m^2_X;t,s,u)$ and $b_{4,5}(m^2_X; s,t,u)=
-b_{4,5}(m^2_X;t,s,u)$. In the following, we will denote the matrix
elements for the decays $X(3872)\to\pi^0\pi^0\chi_{c1}$ and
$X(3872)\to\pi^+\pi^- \chi_{c1} $ as $\mathcal{M}_n$ and
$\mathcal{M}_c$, respectively.

For the rates of the decays $X(3872)\to\pi\pi\chi_{c1}$, the exact
isotopic symmetry predicts the following relation: $\mathcal{B}
(X(3872)\to\pi^0\pi^0\chi_{c1})=\frac{1}{2}\,\mathcal{B}(X(3872)\to
\pi^+\pi^-\chi_{c1})$. As will be shown below, it can be
significantly broken in the real situation.


\section{\boldmath Hadronic loop diagrams for $X(3872)\to\pi\pi\chi_{c1}$}

\begin{figure}  
\begin{center}\includegraphics[width=11cm]{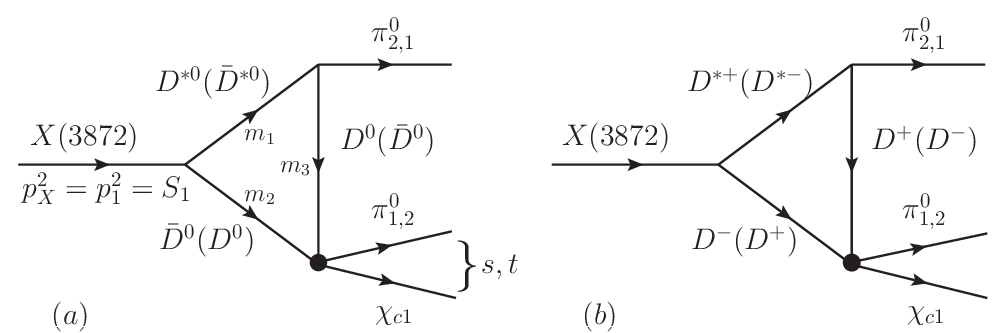}
\caption{\label{Fig1} Eight triangle loop diagrams for the
transition $X(3872)\to\pi^0\pi^0\chi_{c1}$. (a), as well as (b),
involves four diagrams taking into account two charge-conjugate
states in the loops ($D^*\bar DD$ and $\bar D^* D\bar D$) and the
permutation of identical $\pi^0$ mesons.}
\end{center}\end{figure}

\begin{figure}  
\begin{center}\includegraphics[width=11cm]{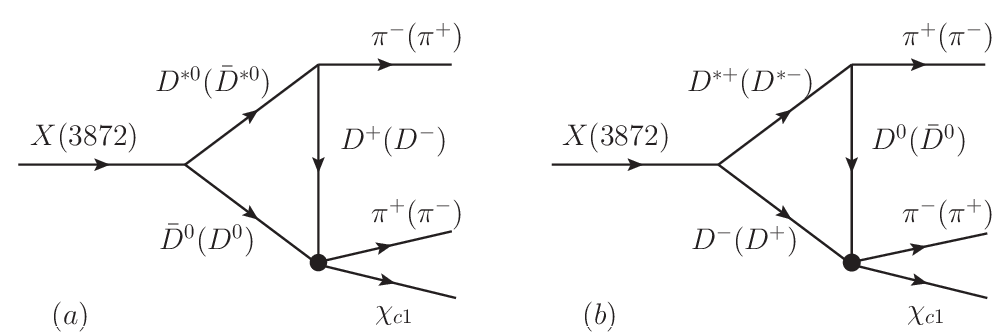}
\caption{\label{Fig2} Four triangle loop diagrams for the transition
$X(3872)\to\pi^+\pi^-\chi_{c1}$. (a), as well as (b), involves two
diagrams with charge-conjugate states in the loops ($D^*\bar DD$ and
$\bar D^* D\bar D$).}
\end{center} \end{figure}

Currently, the mechanism of triangle loop diagrams with charmed
mesons in the loops is considered as a main one of the two-body
decay of $X(3872)\to\pi^0\chi_{c1}$, see in this regard Refs.
\cite{DV08,AS24,FM08,FM12,Gu11,Me15,Zh19,Wu21} and references
herein. We assume that in the three-body decay $X(3872)\to\pi\pi
\chi_{c1}$ the final $\pi\chi_{c1}$ system is produced mainly in a
lower partial wave. This is quite natural in the region near the
$\pi\chi_{c1}$ threshold. Then, the decay of $X(3872)\to\pi\pi
\chi_{c1}$ can be considered as a quasi-two-body process and applied
to its description the mechanism of the triangle loop diagrams.
Examples of such diagrams are shown in Figs. \ref{Fig1} and
\ref{Fig2}. These diagrams (not all) contain so-called triangle
logarithmic singularities \cite{KSW58,Lan59,LT62,FN64,Val64, CN65}.
The literature is rich in examples showing that such singularities
lead to various enhancements in two-body and three-body mass spectra
in the decays of resonances, see, for example, Refs. \cite{LT62,
Val64,Wu12,Ac12,Wu13,AKS15,Ba16,AS19,DZ19,Guo20,Wu24, Xi24} and
references herein.

\begin{figure} 
\begin{center}\includegraphics[width=13cm]{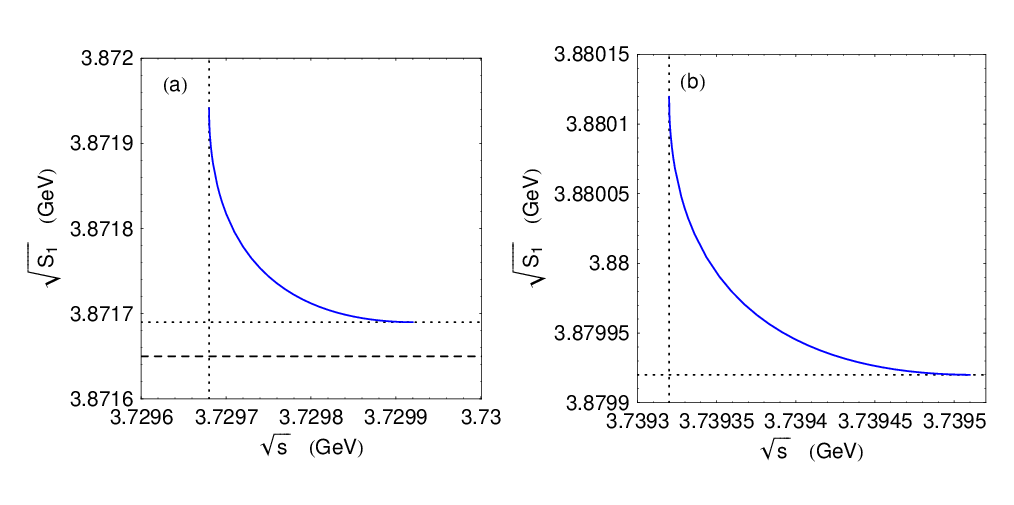}
\caption{\label{Fig3} Solid curves in (a) and (b) show loci of
logarithmic singularities in the $(\sqrt{s},\sqrt{S_1})$ plain for
(a) and (b) in Fig. \ref{Fig1}, respectively. The singularities are
located in (a) in the intervals 3.87169 GeV $<\sqrt{ S_1}<$ 3.87 194
GeV and 3.72968 GeV $<\sqrt{s}<$ 3.72992 GeV, and in (b) in the
intervals 3.87992 GeV $<\sqrt{S_1}<$ 3.88012 GeV and 3.73932 GeV
$<\sqrt{s}<$ 3.73951 GeV.}
\end{center} \end{figure}

The logarithmic singularities in Figs. \ref{Fig1}(a) and
\ref{Fig1}(b) lie along the solid curves shown in Figs.
\ref{Fig3}(a) and \ref{Fig3}(b), respectively. The dependences of
$S_1$ on $s$ given by these curves follow from the equation
$2x_1x_2x_3+x^2_1+x^2_2+x^2_3-1=0$ \cite{Lan59,Val64,FN64, Guo20},
where $x_1=(S_1-m^2_1-m^2_2)/(2m_1m_2)$, $x_2=(s- m^2_2- m^2_3)/(2m_
2m_3)$, and $x_3=(m^2_\pi-m^2_1-m^2_3)/(2m^2_1m^2_3)$, in the
solution of which it is necessary to substitute specific values of
the masses ($m_1$, $m_2$, $m_3$) of particles in the loops [see
notations in Fig. \ref{Fig1}(a)] and the mass of the outgoing $\pi$
meson. At singularity points, all three particles in the loops
simultaneously are on the mass shell \cite{KSW58, Lan59,Val64,
CN65,Guo20}. Of course, this requires that at least one of the
particles corresponding to the internal lines of the diagram is
unstable \cite{Val64,CN65,Guo20}. Horizontal and vertical dotted
lines in Fig. \ref{Fig3}(a) mark the thresholds for the $\sqrt{S_1}$
and $\sqrt{s}$ variables (i.e., the values of $\sqrt{S_1}=m_{D^{*0}
}+ m_{D^0}= 3.87169$ GeV and $\sqrt{s}=2m_{D^0} =3.72968$ GeV) above
which the matrix element $\mathcal{M}_n= \mathcal{M }_n(S_1;s,t,u)$
(see Sec. II) corresponding to Fig. \ref{Fig1}(a) has the imaginary
parts on the $S_1$ and $s$ (or $t$) variables. Intervals containing
the curve of singularities, $m_1+m_2<\sqrt{S_1} <\sqrt{ m^2_1+
m^2_2+m_2m_3+m_2( m^2_1-m^2_\pi)/m_3}$ and $m_2+m_3
<\sqrt{s}<\sqrt{m^2_2+m^2_3+m_ 1m_2+m_2 (m^2_3-m^2_\pi)/m_1}$, are
bounded by the points, where this curve touches the above lines
(see, for example, Ref. \cite{Guo20}). The horizontal dashed line in
Fig. \ref{Fig3}(a) marks the nominal mass of the $X(3872)$ state
$m_X=3.87165$ GeV \cite{PDG23}. Since the width of the $X(3872)$,
$\Gamma_X$, is not less than 1 MeV \cite{PDG23,Aai20,Hir23,Abl23},
and the available values of $\sqrt{s}$ lie in the range from
$m_{\chi_{1c}}+m_{ \pi^0}=3.64565$ GeV to $m_X-m_{\pi^0}=3.73667$
GeV, then the locus of logarithmic singularities of triangle in Fig.
\ref{Fig1}(a) completely falls into the physical region of the
$X(3872)\to\pi^0\pi^0\chi_{c1}$ decay.

Let us move on to Fig. \ref{Fig3}(b) associated with  in Fig.
\ref{Fig1}(b). The threshold values of $\sqrt{S_1}=m_{D^{*+}}+m_{
D^-}=3.87992$ GeV and $\sqrt{s}=2m_{D^\pm}=3.73932$ GeV marked by
horizontal and vertical dotted lines lie 8.23 and 9.64 MeV above the
thresholds of the $D^{*0}\bar D^0$ and $D^{0}\bar D^0$ channels,
respectively. It is clear that the triangle singularities of the
diagrams with charged charmed $D^*$ and $D$ mesons in the loops are
located outside the physical region of the $X(3872)\to\pi^0\pi^0
\chi_{c1}$ decay. However, the contribution of Fig. \ref{Fig1}(b),
as will be shown in the next section, turns out to be important and
must be taken into account.

Let us now consider the diagrams in Fig. \ref{Fig2} for the decay
$X(3872)\to\pi^+\pi^-\chi_{c1}$. In Fig. \ref{Fig2}(a), there are no
triangle singularities, since the decay channel of the $D^{*0}$ into
$\pi^- D^+$ is closed ($m_{D^{*0}}=2.00685$ GeV, $m_{D^+}+m_{\pi^-}=
2.00923$ GeV, and $\Gamma_{D^{*0}}\simeq55.6$ keV \cite{AS19}).
Figure \ref{Fig2}(b) have triangle singularities. But they lie in
the region of $3.87992\,\mbox{GeV}\,<\sqrt{S_1}<
3.88014\,\mbox{GeV}$ and $3.7345\,\mbox{GeV}\,<\sqrt{s}<
3.73471\,\mbox{GeV}$ which on the $\sqrt{S_1}$ variable starts 8.27
MeV above the nominal mass of the $X(3872)$ and on the $\sqrt{s}$
variable 2.42 MeV to the right of the maximum permissible value of
$\sqrt{s}=m_X-m_{\pi^-}=3.73208$ GeV in this decay. The values
$\sqrt{S_1}=3.87992$ GeV and $\sqrt{s} =3.7345$ GeV indicate the
thresholds of the $D^{*+}D^-$ and $D^0D^-$ channels, respectively.
Thus, both of these channels are closed in the
$X(3872)\to\pi^+\pi^-\chi_{c1}$ decay and the amplitude for Fig.
\ref{Fig2}(b) turns out to be purely real (if neglect by the tiny
value of $\Gamma_{D^{*+}}$ in the $D^{*+}$ meson propagator). How
the contributions of Figs. \ref{Fig2}(a) and \ref{Fig2}(b)
correlate to each other, we will find out in the next section.


\section{\boldmath Estimates of $\mathcal{B}(X(3872)\to\pi\pi\chi_{c1})$}


To estimate $\mathcal{B}(X(3872)\to\pi\pi\chi_{c1})$ we restrict
ourselves to the contributions of the diagrams presented in Figs.
\ref{Fig1} and \ref{Fig2}. First of all, consider the amplitude of
the subprocess $D\bar D\to\pi\chi_{c1}$ which is a component part of
the matrix element $\mathcal{M}$. We will estimate it on the mass
shell near the $D\bar D$ threshold and then use the found value as
an effective ``coupling constant'' characterizing the $D\bar
D\pi\chi_{c1}$ vertex in the triangular loops. The isotopic
invariance of strong interactions and the $P$-parity conservation
allow us to write down a number of of useful relations for the
reaction $D\bar D\to\pi\chi_{c1}$:
\begin{eqnarray}\label{Eq6}
I_f=1=I_i,\ \ G_f=-1=G_i=(-1)^{I_i+l_i},\ \ l_i=0,2,...,\ \ J=l_i,\
\ P_i=(-1)^{l_i}=P_f=-(-1)^{l_f},\ \ l_f=1,3,...,\end{eqnarray}
where the indices $i$ and $f$ indicate the belonging of quantum
numbers to the initial $D\bar D$ and final $\pi\chi_{c1}$ states,
respectively; $I$, $G$, $l$, $J$, and $P$ are the isospin, $G$
parity, orbital moment, total moment, and $P$ parity, respectively.
For $l_i=0$ ($J=0$) there is only one possible value of $l_f=1$, and
for each $l_i\geq2$ two values $l_f=l_i\pm1$ are allowed. The
partial amplitude of the process $D\bar D\to\pi \chi_{c1}$ with
$l_i=0$ ($J=0$) and $l_f=1$ experiences a minimal suppression caused
by the threshold factors near the threshold. This amplitude has the
form
\begin{eqnarray}\label{Eq7} f^{J=0}_{D\bar D\pi\chi_{c1}}=g_{D\bar
D\pi \chi_{c1}}\left(\vec{p}_{\chi_{c1}}(s),\vec{\xi}^*\right),
\end{eqnarray} where $\vec{p}_{\chi_{c1}}(s)$ is the momentum of
the $\chi_{ c1}$ meson in the $D\bar D$ center-of-mass system, and
$\vec{\xi}$ is the polarization vector of the $\chi_{c1}$ in its
rest frame (see Ref. \cite{FN1}); $|\vec{p}_{\chi_{c1}}(s)|=\sqrt{
s^2-2s(m^2_{\chi_{c1}}+m^2_\pi)+ (m^2_{\chi_{c1}}-m^2_\pi)^2}/(2
\sqrt{s})$. It is quite natural to assume that the factor $g_{D\bar
D\pi\chi_{c1}}$ near the threshold is a smooth function of
$\sqrt{s}$. We will calculate it for $\sqrt{s}=2m_D$ assuming that
the reaction $D\bar D\to\pi\chi_{c1}$ (near the threshold) proceeds
via $D^*$ exchanges in its $t$ and $u$ channels. In this simple
model we have
\begin{eqnarray}\label{Eq8}
g_{D\bar D\pi\chi_{c1}}=g_{D^*D\pi}\,g_{\chi_{c1}D^*\bar D}\,\frac{4
\,m_D }{ m_{\chi_{c1}}}\,\frac{3+m^2_D/m^2_{D^*}}{2m^2_D+ 2m^2_{D^*}
-m^2_{ \chi_{c1}}}=g_{D^*D\pi}\,g_{\chi_{c1}D^*\bar D}\times
(3.05696\ \mbox{GeV}^{-2}), \end{eqnarray} where $g_{D^*D\pi}$ and
$g_{\chi_{c1}D^*\bar D}$ are the coupling constants in the
interaction vertices $V_{D^*D\pi}=g_{D^*D\pi}(\varepsilon^*_{
D^*},p_\pi+p_D)$ and $V_{\chi_{c1}D^*\bar D}=g_{\chi_{c1}D^*\bar
D}(\varepsilon_{D^*}, \varepsilon^*_{\chi_{c1}})$. When obtaining
Eq. (\ref{Eq8}), we neglected the mass squared of the $\pi$ meson,
and also put $m_{D^{*+}}=m_{D^{*0}}$ and $m_{D^+ }=m_{D^0}$. At the
$D\bar D$ threshold, the virtuality of the exchanged $D^*$ mesons
(i.e., $m^2_{D^*}-q^2$, where $q$ is the four-momentum of the $D^*$)
is approximately $1.34 3$ GeV$^2$. In order to take into account to
some extent the internal structure and the off-mass-shell effect for
the  $D^*$ meson, it is necessary to introduce the form factor  into
the each vertices of the $D^*$ exchange:
$\mathcal{F}(q^2,m^2_{D^*})=\frac{\Lambda^2-m^2_{ D^*}}{\Lambda^2
-q^2}$ \cite{Go96,Co02,Co04,Ch05,AS24}. Here we orient on the
typical value of the parameter $\alpha\approx2$ \cite{AS24}
associated with the $\Lambda$ by the relation $\Lambda=m_{D^*}+
\alpha\Lambda_{\scriptsize \mbox{QCD}}$ \cite{Ch05}, where
$\Lambda_{\scriptsize\mbox{QCD}}=220$ MeV. This form factor results
in decreasing the effective coupling constant $g_{D\bar D\pi\chi_{
c1}}$ by approximately 2.84 times in comparison with the estimate in
Eq. (\ref{Eq8}); $g^2_{D\bar D\pi \chi_{c1}}$ decreases by a factor
of 8.06 accordingly. Next we will use for $g_{D\bar D\pi\chi_{c1}}$
the value of $g_{D^*D\pi}\,g_{ \chi_{c1}D^*\bar D}\times(1.07647\
\mbox{GeV}^{-2})$ obtained taking into account the form factor. From
the isotopic symmetry for the coupling constants $g_{D^*D\pi}$ and
the data on the decays $D^{*+}\to(D\pi)^+$ \cite{PDG23}, it follows
that $g_{D^{*0}D^0\pi^0} =g_{D^{*0}D^+\pi^-}/\sqrt{2}=
g_{D^{*+}D^0\pi^+}/\sqrt{2}=-g_{D^{*+ }D^+\pi^0}\approx5.93$
\cite{AS19}. The constant $g_{\chi_{c1}D^* \bar D}$ cannot be
measured directly, but its value is predicted theoretically within
the framework of the effective theory of heavy quarks
\cite{Wu21,Co02,Co04,Ch05,Gu11,Me07}: $g_{\chi_{c1 }D^*\bar
D}=2\sqrt{2}g_1\sqrt{m_Dm_{D^*}m_{\chi_{c1}}}=(-21.45\pm1.68)$ GeV
\cite{AS19}, where $g_1$ is an universal constant. As a result, we
get $g_{D^0\bar D^0\pi^0\chi_{c1}}=-g_{D^+D^-\pi^0 \chi_{c1}}=g_{D^+
\bar D^0\pi^+\chi_{c1}} /\sqrt{2}=g_{D^0D^-\pi^-\chi_{c1}}/\sqrt{2}
\approx137$ GeV$^{-1}$ and will use this value as a guide.

The above structure of the $D\bar D\pi\chi_{c1}$ vertex allows us to
write the matrix element $\mathcal{M}_n(S_1;s,t,u)$ for the
contribution of the eight diagrams in Fig. \ref{Fig1} as follows:
\begin{eqnarray}\label{Eq9}
\mathcal{M}_n(S_1;s,t,u)=2\frac{\bar{g}}{16\pi}\varepsilon^\mu_X
\left[I_\mu(p_1,p_4)\left(\vec{p}_{\chi_{c1}}(s),\vec{\xi}^*\right)+
I_\mu(p_1,p_3)\left(\vec{p}_{\chi_{c1}}(t),\vec{\xi}^*\right)\right.
\nonumber\\ \left.+ \tilde{I}_\mu(p_1,p_4)\left(\vec{p}_{\chi_{c1}}
(s),\vec{\xi}^*\right) +\tilde{I}_\mu(p_1,p_3)\left(\vec{p}_{
\chi_{c1}}(t),\vec{\xi}^*\right) \right],
\end{eqnarray}
where the common factor 2 arises owing to the equality of the
contributions from the loops with the charge conjugated intermediate
states, $\bar{g}=g_Xg_{D^{*0}D^0\pi^0}g_{D^0\bar D^0 \pi^0
\chi_{c1}}$, $g_X$ is the coupling constant of the $X(3872)$ to
$D^{*0}\bar D^0$ in the vertex $V_{XD^{*0}\bar D^0}=g_X
(\varepsilon_X,\varepsilon^*_{D^{*0}})$ (the values of $g_X$ will be
specified below); the amplitude $I_\mu(p_1,p_4)$ represents the
following vector integral
\begin{eqnarray} \label{Eq10} I_\mu(p_1,p_4)=\frac{i}{\pi^3}\int
\frac{\left(-g_{\mu\nu}+\frac{k_\mu k_\nu}{ m^2_{D^{*0}}}\right)
(2p_{4\nu}-k_\nu)\,d^4k}{(k^2- m^2_{D^{*0}}+i\epsilon)((p_1-k)^2
-m^2_{\bar D^0}+i\epsilon)((k-p_4)^2-m^2_{ D^0}+i\epsilon)}.
\end{eqnarray} The amplitude $I_\mu(p_1,p_3)$, $\tilde{I}_\mu(p_1,
p_4)$, and $\tilde{I}_\mu(p_1,p_3)$ have a similar form. In so
doing,  $I_\mu (p_1, p_4)$ and $I_\mu(p_1,p_3)$ correspond to Fig.
\ref{Fig1}(a) which differ in the permutation of identical $\pi^0$
mesons, and $ \tilde{I}_\mu(p_1,p_4)$ and $\tilde{I}_\mu(p_1,p_3)$
correspond to similar Fig. \ref{Fig1}(b) in the same figure. In Ref.
\cite{AS19} it was shown that the divergent part of a vector
integral of type (\ref{Eq10}) is proportional to $p_{1\mu}$ [i.e.,
the four-momentum of the $X(3872)$ resonance] and it does not
contribute to the matrix element $\mathcal{M}_n( S_1;s,t, u)$
because $(\varepsilon_X,p_1)=0$. It was also shown in Ref.
\cite{AS19} that its convergent part, $I^{\scriptsize{
\mbox{conv}}}_\mu( p_1,p_4)$, proportional to $p_{4\mu}$ is
dominated by the amplitude of the scalar triangle diagram, which we
denote here as $I(S_1, s)$, i.e.,
$I^{\scriptsize{\mbox{conv}}}_\mu(p_1, p_4)=-2p_{4\mu}I(S_1,s)$,
where
\begin{eqnarray} \label{Eq11} I(S_1,s)=\frac{i}{\pi^3}\int\frac{d^4k}{(k^2-
m^2_{D^{*0}}+i \epsilon)((p_1-k)^2-m^2_{\bar D^0}+i\epsilon)(
(k-p_4)^2-m^2_{ D^0}+i\epsilon)}. \end{eqnarray} As a result, Eq.
(\ref{Eq9}) takes the form
\begin{eqnarray}\label{Eq12}
\mathcal{M}_n(S_1;s,t,u)=-4\frac{\bar{g}}{16\pi}\left\{(\varepsilon_X,
p_4)[I(S_1,s)+\tilde{I}(S_1,s)]\left(\vec{p}_{\chi_{c1}}(s),\vec{\xi}^*
\right)+(\varepsilon_X,p_3)[I(S_1,t)+\tilde{I}(S_1,t)]\left(\vec{p}_{
\chi_{c1}}(t),\vec{\xi}^* \right)\right\}. \end{eqnarray} About the
contributions of the scalar amplitudes $I(S_1,s)$ and $\tilde{I}
(S_1,s)$ one can speak as of the $s$ contributions from Fig.
\ref{Fig1}(a) and \ref{Fig1}(b), respectively, and about the
contributions of the scalar amplitudes $I(S_1,t)$ and $\tilde{I}(
S_1,t)$ one can speak as of the $t$ contributions from Fig.
\ref{Fig1}(a) and \ref{Fig1}(b) with permutation of identical
$\pi^0$ mesons, respectively.

To numerically calculate scalar triangle amplitudes, we use explicit
formulas obtained in Refs. \cite{Guo20,Gu11} within the framework of
nonrelativistic formalism. We convinced that the results of such a
calculation are in excellent agreement with what is given for these
amplitudes the exact expressions through dilogarithms \cite{Den07}.
We take into account the finite width of the $D^{*0} $ meson by
replacing $m^2_{D^{*0}}$ in its propagator with $m^2_{D^{
*0}}-im_{D^ {*0}}\Gamma_{D^{*0}}$ and put $\Gamma_{D^{*0}}= 55.6$
keV \cite{AS19}. This leads to a significant smoothing and reduction
in the contributions of  triangle logarithmic  singularities to
$\mathcal{M}_n(S_1;s,t,u)$ as compared with the hypothetical case
corresponding to $\Gamma_{D^{*0}}=0$. The finite width of the
$D^{*+}$ meson, $\Gamma_{D^{*+}}=83.6$ keV, is taken into account in
a similar way.

\begin{figure} 
\begin{center}\includegraphics[width=13cm]{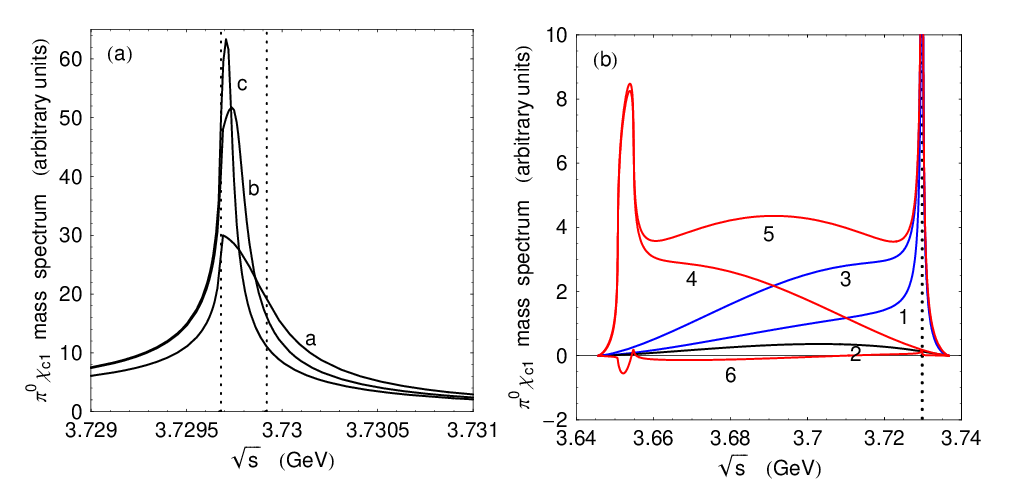}
\caption{\label{Fig4} (a) The solid curves a, b, and c show the
examples of the $\pi^0\chi_{c1}$ mass spectrum $d\Gamma(X(3872)
\to\pi^0 \pi^0\chi_{c1}; S_1,s)/d\sqrt{s}$ in the region of the
$D^0\bar D^0$ threshold calculated using Eq. (\ref{Eq14}) at
$\sqrt{S_1}=3.87165$, 3.87172, and 3.87177 GeV, respectively. The
the dotted vertical lines mark the $\sqrt{s}$ values between which
the amplitude of the $X(3872)\to(D^{*0}\bar D^0+\bar D^{*0}D^0)
\to\pi^0\pi^0\chi_{c1}$ decay contains the logarithmic singularities
which manifest themselves in the $\pi^0 \chi_{c1}$ mass spectrum as
narrow peaks. (b) The components of the $\pi^0\chi_{c1}$ mass
spectrum at $\sqrt{S_1}= 3.87172$ GeV throughout the accessible
region of $\sqrt{s}$; description of the curves see in the text.
}\end{center}\end{figure}

The differential probability of the $X(3872)\to\pi^0\pi^0\chi_{c1}$
decay which determines the distribution of events in the Dalitz plot
has the form \cite{PDG23}:
\begin{eqnarray}\label{Eq13} \frac{d^2\Gamma(X(3872)\to\pi^0\pi^0\chi_{
c1};S_1;s,t,u)}{dtds}=\frac{1}{3(2\pi)^332S_1^{3/2}}\sum_{\lambda\lambda'}
\left|\mathcal{M}_n(S_1;s,t,u)\right|^2, \end{eqnarray} where
summation over $\lambda$ and $\lambda'$ means summation over
polarizations of the $X(3872)$ and $\chi_{c1}$ mesons, respectively.
We write the mass spectrum of the $\pi^0\chi_{c1}$ system over the
$\sqrt{s}$ variable as
\begin{eqnarray}\label{Eq14}
\frac{d\Gamma(X(3872)\to\pi^0\pi^0\chi_{c1};S_1,s)}{d\sqrt{s}}=2\sqrt{s}
\int\limits_{t_-(S_1,s)}^{t_+(S_1,s)}\frac{d^2\Gamma(X(3872)
\to\pi^0\pi^0\chi_{c1};S_1;s,t,u)}{dtds}\,dt,\end{eqnarray} where
$t_{\pm}(S_1,s)$ denote the boundaries of the physical region for
the $t$ variable for fixed values of $s$ and $S_1$ \cite{PDG23}.
Figure \ref{Fig4}(a) shows examples of the $\pi^0\chi_{c1}$
unnormalized mass spectra near the $D^0\bar D^0$ threshold for
several values of $\sqrt{S_1}$. These examples illustrate the
resonantlike manifestations of the triangle singularities present in
the amplitude $I(S_1,s)$. Figure \ref{Fig4}(b) shows [in the same
units as in Fig. \ref{Fig4}(a)] all significant components of the
$\pi^0\chi_{c1}$ mass spectrum at $\sqrt{S_1}=3.87172$ GeV
throughout the accessible region of $\sqrt{s}$. Curves 1, 2, and 3
correspond to the contributions of the amplitudes $I(S_1,s)$ [Fig.
\ref{Fig1}(a)], $\tilde{I}(S_1,s)$ [Fig. \ref{Fig1}(b)], and their
sum $I(S_1,s)+\tilde{I}(S_1,s) $, respectively. Curve 4 corresponds
to the contribution of the amplitude $I(S_1,t)+\tilde{I}(S_1,t)$
[from the sum of Fig. \ref{Fig1}(a) and \ref{Fig1}(b)with the
transposed identical $\pi^0$ mesons]. The contributions of the
amplitudes $I(S_1,t)$ and $\tilde{I}(S_1,t)$ are not shown
separately so as not to clutter the figure. Curve 6 corresponds to
the contribution of interference between the amplitudes
$I(S_1,s)+\tilde{I}(S_1,s)$ and $I(S_1,t)+\tilde{I}(S_1,t)$ which
differ by permutation of identical $\pi^0$ mesons. It can be seen
that the interference is small for all values of $\sqrt{s}$ and can
be neglected. The total contribution to the $\pi^0\chi_{c1}$ mass
spectrum from the amplitudes $(I(S_1,s)+\tilde{I}(S_1,s))$ and
$(I(S_1,t)+\tilde{I} (S_1,t))$ in neglecting their interference is
shown in Fig. \ref{Fig4}(b) by curve 5. If the peak in the
$\pi^0\chi_{c1}$ mass spectrum over the $\sqrt{s}$ variable in the
vicinity of $\sqrt{s} \approx2m_{D^0} \approx3.72968$ GeV is due to
triangle singularities in the amplitude $I(S_1,s)$, then the peak in
the region $3.65\ \mbox{GeV} <\sqrt{s}<3.6575$ GeV is a
manifestation in the distribution over $\sqrt{s}$ of the triangle
singularities in the amplitude $I(S_1,t)$.

\begin{figure} 
\begin{center}\includegraphics[width=13cm]{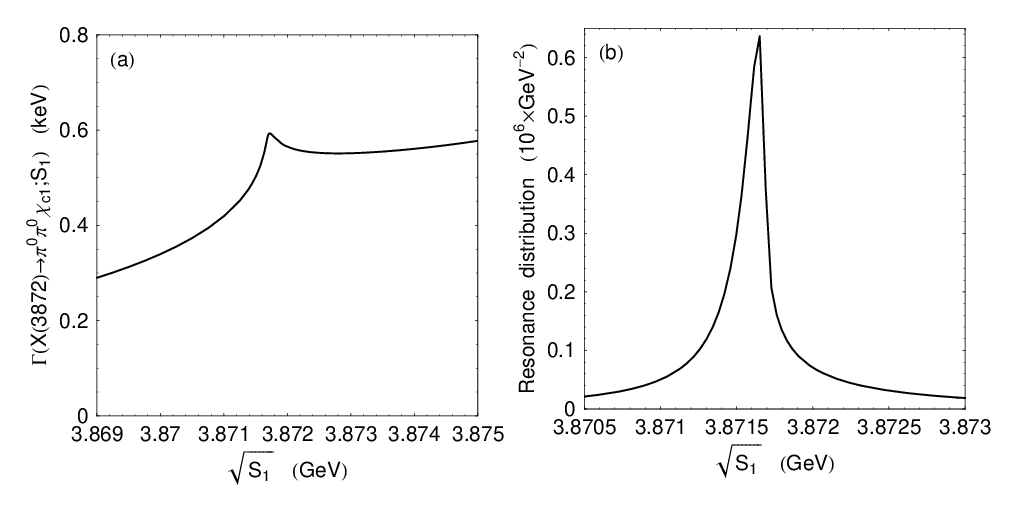}
\caption{\label{Fig5} (a) The solid curve shows the width $\Gamma
(X(3872)\to\pi^0\pi^0\chi_{c1};S_1)$ calculated  using  Eq.
(\ref{Eq15}). The constructed example corresponds to
$g^2_X/(16\pi)=0.25$ GeV$^2$. (b) An example of the resonance
distribution $2S_1/(\pi| D_X(S_1)|^2)$ for the $X(3872)$ at
$g^2_X/(16\pi)=0.25$ GeV$^2$ and $\Gamma_{\scriptsize\mbox{non}}=1$
MeV \cite{AS19}.}\end{center}\end{figure}

The width of the $X(3872)\to\pi^0\pi^0\chi_{c1}$ decay in the
general case is determined by the expression
\begin{eqnarray}\label{Eq15}
\Gamma(X(3872)\to\pi^0\pi^0\chi_{c1};S_1)=\frac{1}{2} \int\limits_{
(m_{\chi_{c1}}+m_{\pi^0})^2}^{(\sqrt{S_1}-m_{\pi^0})^2} ds\int
\limits_{t_-(S_1,s)}^{t_+(S_1,s)}\frac{d^2\Gamma(X(3872) \to\pi^0
\pi^0\chi_{c1};S_1;s,t,u)}{dtds}\,dt,\end{eqnarray} where the factor
$1/2$ takes into account the identity of $\pi^0$ mesons. In Fig.
\ref{Fig5}(a), we presented the result of the calculation of
$\Gamma(X(3872)\to\pi^0\pi^0\chi_{c1};S_1)$ using the coupling
constant $g^2_X/( 16\pi)=0.25$ GeV$^2$ as a guide (see
\cite{AR14,AS19,AS24}). The maximum of the width $\Gamma(X(3872)
\to\pi^0\pi^0\chi_{c1};S_1)$ near the $D^{*0}D^0$ threshold  is
caused by the presence in the amplitude of the triangle
singularities.

To estimate $\mathcal{B}(X(3872)\to\pi^0\pi^0\chi_{c1})$ it is
necessary to weigh the energy dependent width $\Gamma(X(3872)
\to\pi^0 \pi^0\chi_{c1};S_1)$ with the resonance distribution
$2S_1/(\pi|D_X(S_1)|^2)$:
\begin{eqnarray}\label{Eq16}
\mathcal{B}(X(3872)\to\pi^0\pi^0\chi_{c1})=\int\limits^{\infty}_{
m_{\chi_{c1}}+2m_{\pi^0}}\frac{2\sqrt{S_1}}{\pi}\frac{\sqrt{S_1}
\Gamma(X(3872)\to\pi^0 \pi^0\chi_{c1};S_1)}{|D_X(S_1)|^2}
\,d\sqrt{S_1},\end{eqnarray} where $D_X(S_1)$ is the inverse
propagator of the $X(3872)$ which we take from Refs. \cite{AR14,
AS19}. Note that the resonance distribution $2S_1/(\pi|
D_X(S_1)|^2)$ has good analytical and unitary properties
\cite{AR14,AS19}. Figure \ref{Fig5}(b) shows an example of this
distribution calculated at $m_X=3871.65$ MeV, $g^2_X/(16\pi)=0.25$
GeV$^2$, and $\Gamma_{\scriptsize\mbox{non}}=1$ MeV, where
$\Gamma_{\scriptsize \mbox{non}}$ approximately describes the width
of the $X(3872)$ decay into all non-$(D^*\bar D+\bar D^*D)$
channels. Of course, the main contribution to the integral
(\ref{Eq16}) comes from the narrow region of the resonance peak. The
result of integration over the region $3.869\ \mbox{GeV}<\sqrt{S_1}
<3.875$ GeV for the above parameter values gives $\mathcal{B}
(X(3872)\to\pi^0\pi^0\chi_{c1})\approx 1.24\times10^{-4}$.
Table I shows the estimates of $\mathcal{B}(X(3872)\to\pi^0\pi^0
\chi_{c1})$ for different values of $g^2_X/(16\pi)$ and $\Gamma_{
\scriptsize\mbox{non}}$ which we vary in a reasonable range taking
into account the current (far from final) information about the
$X(3872)$ obtained from the analyses of its main decay channels in
Refs. \cite{AR14,AS19,AS24,Aai20,Hir23,Abl23}.

\begin{table} 
\centering \caption{$\mathcal{B}(X(3872)\to\pi\pi\chi_{c1})$ in
units of $10^{-4}$.}\label{Tab1}\vspace*{0.1cm}
\begin{tabular}{l c c c c c c }
 \hline\hline
 $g^2_X/(16\pi)$ (in GeV$^2$)     \ \  &\ \  0.25 &\ \  0.5   &\ \  0.671 &\ \ \ \ 0.25  &\ \  0.5   &\ \  0.671\\
 $\Gamma_{\scriptsize\mbox{non}}$\ \  &\ \       &\ \  1 MeV &\ \        &\ \        &\ \  2 MeV &\ \  \\ \hline
 $\mathcal{B}(X(3872)\to\pi^0\pi^0\chi_{c1})$\ \  &\ \  1.24 &\ \  1.63 &\ \  1.61  &\ \ \ \ 0.77  &\ \  0.88  &\ \  0.90 \\
 $\mathcal{B}(X(3872)\to\pi^+\pi^-\chi_{c1})$\ \  &\ \  1.51 &\ \  1.77 &\ \  1.73  &\ \ \ \ 0.86  &\ \  0.97  &\ \  0.99\\ \hline
 $\mathcal{R}=\frac{\mathcal{B}(X(3872)\to\pi^+\pi^-\chi_{c1})}{\mathcal{B}(X(3872)\to\pi^0\pi^0\chi_{c1})}$
 &\ \   1.22 &\ \   1.09 &\ \   1.07 &\ \ \ \  1.12 &\ \   1.10 &\ \   1.10 \\
 \hline\hline
\end{tabular} \end{table}

\begin{figure} [!ht] 
\begin{center}\includegraphics[width=13cm]{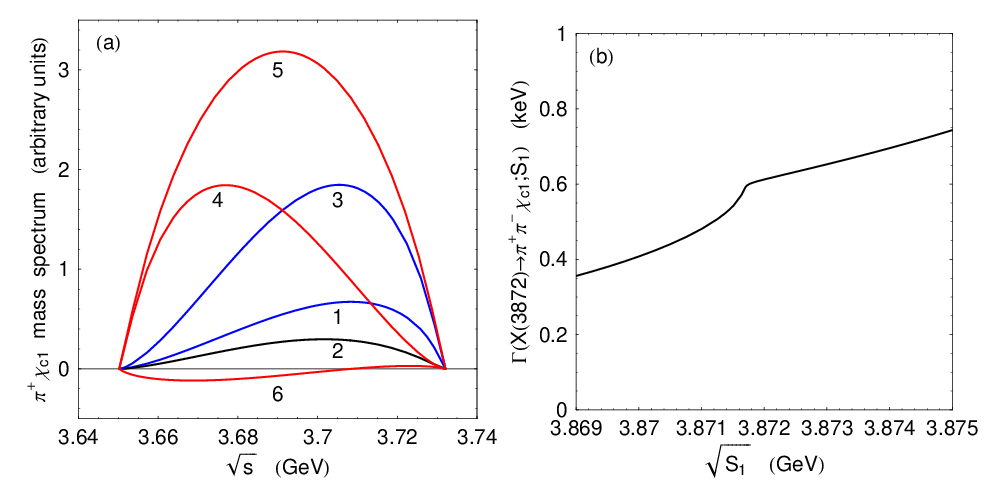}
\caption{\label{Fig6} (a) The components of the $\pi^+\chi_{c1}$
mass spectrum at $\sqrt{S_1}=3.87172$ GeV throughout the accessible
region of $\sqrt{s}$ in the same units as in Fig. \ref{Fig4}; the
curves here have the same meaning as the curves with the
corresponding numbers in Fig. \ref{Fig4}(b) which have been
described in detail above in the text. (b) The width
$\Gamma(X(3872)\to\pi^+\pi^- \chi_{c1};S_1)$ as a function of
$\sqrt{S_1}$. The constructed example corresponds to
$g^2_X/(16\pi)=0.25$ GeV$^2$.}\end{center}\end{figure}

Let us now consider the diagrams in Fig. \ref{Fig2} describing the
decay of $X(3872)\to\pi^+\pi^-\chi_{c1}$. Although there are only
four of such diagrams, and not eight as in Fig. \ref{Fig1}, the
factor of 2 in Eq. (\ref{Eq9}) is preserved also for the amplitude
$\mathcal{M}_c(S_1;s,t,u)$ owing to the isotopic factors in the
$D^*D\pi$ and $D\bar D\pi\chi_{c1}$ vertices, which are indicated
above in the paragraph after Eq. (\ref{Eq8}). Thus, with taking into
account the replacement of the particle masses in the loops and the
masses of the final pions, as well as the necessary changes in
designations and exclusion of the factor $1/2$ from Eq. (\ref{Eq15})
when determining the width of the $X(3872)\to\pi^+\pi^-\chi_{c1}$
decay, we can use Eqs. (\ref{Eq9})--(\ref{Eq16}) to calculate the
$\pi^\pm\chi_{c1}$  mass spectra, $\Gamma(X(3872)
\to\pi^+\pi^-\chi_{c1};S_1)$, and $\mathcal{B}(X(3872)\to\pi^+ \pi^
-\chi_{c1})$. Figure \ref{Fig6}(a) shows (in the same units as in
Fig. \ref{Fig4}) the main components of the $\pi^+\chi_{c1}$ mass
spectrum in the $X(3872)\to\pi^+\pi^-\chi_{c1}$ decay at $\sqrt{
S_1}= 3.87172$ GeV throughout the accessible region of $\sqrt{s}$.
The curves here have the same meaning as the curves with the
corresponding numbers in Fig. \ref{Fig4}(b), which have been
described in detail above in the text. In this case, there are no
triangle singularities in the physical region of the decay  and the
$\pi^\pm \chi_{c1}$ mass spectra  are smooth functions of
$\sqrt{s}$. The energy dependent decay width $\Gamma(X(3872)\to
\pi^+\pi^-\chi_{c1};S_1)$ [see the example shown in Fig.
\ref{Fig6}(b)] has a characteristic break at the threshold of the
$D^{*0}\bar D^0$ channel.

The estimates for $\mathcal{B}(X(3872)\to\pi^+\pi^-\chi_{c1})$ are
given in Table I. We see that the model under discussion predicts
the close values for $\mathcal{B}(X(3872)\to\pi^+ \pi^-\chi_{c1})$
and $\mathcal{B}(X(3872)\to\pi^0\pi^ 0 \chi_{c1})$ the absolute
values of which turn out to be at the level of about $10^{-4}$.
Their ratio averaged over the variants in Table I, $\mathcal{R}=
\mathcal{B}(X(3872)\to\pi^+\pi^-\chi_{c1})/\mathcal{B}(X(3872)\to
\pi^0\pi^0\chi_{c1 })\approx1.1$, indicates a noticeable violation
of isotopic symmetry, according to which one would expect
$\mathcal{R} =2$.

\begin{table} [!ht]
\centering \caption{$\mathcal{B}(X(3872)\to\pi\pi\chi_{c1})$ (in
units of $10^{-4}$) only for Figs. \ref{Fig1}(a) and
\ref{Fig2}(b).}\label{Tab1}\vspace*{0.1cm}
\begin{tabular}{l c c c c c c }
 \hline\hline
 $g^2_X/(16\pi)$ (in GeV$^2$)     \ \  &\ \  0.25 &\ \  0.5   &\ \  0.671 &\ \ \ \ 0.25  &\ \  0.5   &\ \  0.671\\
 $\Gamma_{\scriptsize\mbox{non}}$\ \  &\ \       &\ \  1 MeV &\ \        &\ \        &\ \  2 MeV &\ \  \\ \hline
 $\mathcal{B}(X(3872)\to\pi^0\pi^0\chi_{c1})$\ \  &\ \  0.662 &\ \  0.741 &\ \  0.731  &\ \ \ \ 0.338  &\ \  0.393  &\ \  0.404 \\
 $\mathcal{B}(X(3872)\to\pi^+\pi^-\chi_{c1})$\ \  &\ \  0.526 &\ \  0.614 &\ \  0.602  &\ \ \ \ 0.296  &\ \  0.337  &\ \  0.342\\ \hline
 $\mathcal{R}=\frac{\mathcal{B}(X(3872)\to\pi^+\pi^-\chi_{c1})}{\mathcal{B}(X(3872)\to\pi^0\pi^0\chi_{c1})}$
 &\ \   0.795 &\ \   0.829 &\ \   0.824 &\ \ \ \  0.876 &\ \   0.858 &\ \   0.847 \\
 \hline\hline
\end{tabular} \end{table}

The present calculation assumes that the $X(3872)$ is a pure
charmonium, and this is reflected in the equal couplings of
$X(3872)\to D^{*0}\bar D^0$ and $D^{*+}D^-$. In a molecular
interpretation of the $X(3872)$, $X(3872)$ couples differently with
$D^{*0}\bar D^0$ and $D^{*+} D^-$. For example, in Ref. \cite{FM08},
$X(3872)\to D^{*0}\bar D^0$ is considered while $X(3872)\to D^{*+}
D^-$ neglected. In this regard, we present in Table II the values of
the branching fractions corresponding only to Figs. \ref{Fig1}(a)
and \ref{Fig2}(a). In a sense, this corresponds to the limiting
variant of the molecular model when the $X(3872)$ is associated only
with the $D^{*0}\bar D^0+c.c.$ channel.


\section{\boldmath Conclusion}

We have obtained the tentative estimates for $\mathcal{B}(X(3872)
\to\pi^0\pi^0\chi_{c1})$ and $\mathcal{B}(X(3872)\to\pi^+\pi^-
\chi_{c1})$ in the model of the triangle loop diagrams with charmed
$D^*\bar DD$ and $\bar D^*D\bar D$ mesons in the loops. The decay
rates are predicted at the level of $10^{-4}$ at the reasonable
values of the coupling constants. We would like to draw a special
attention to the fact that in this model an important contribution
to $\mathcal{B}(X(3872)\to\pi\pi\chi_{c1})$ is given by the
(``heavy'') charged $D^{*+} D^-+c.c.$ intermediate states,
certainly, together with the (``light'') neutral $D^{*0}\bar
D^0+c.c.$ intermediate states. This is obvious from Figs.
\ref{Fig4}(b) and \ref{Fig6}(a).

Within the framework of the considered model, the decay rates
$X(3872)\to\pi^0\pi^0\chi_{c1}$ and $X(3872)\to\pi^+\pi^-\chi_{c1}$
are proportional to the same product of coupling constants. The
existing uncertainties in these constants, as well as the remaining
(so far) uncertainties in such characteristics of the $X(3872)$
resonance as its mass $m_X$ and width $\Gamma_{\scriptsize
\mbox{non}}$ \cite{PDG23, AS24, Aai20, Hir23,Abl23} allow us only to
hope (before the experiment) that the model correctly predicts the
order of magnitude of the probabilities for the $X(3872)\to\pi^0
\pi^0\chi_{c1}$ and $X (3872)\to\pi^+ \pi^- \chi_{c1}$ decays. The
ratio $\mathcal{R}=\mathcal{B}(X(3872)\to \pi^+ \pi^+\chi_{c1})/
\mathcal{B}(X(3872)\to\pi^0\pi^0\chi_{c1}) \approx1.1$ does not
depend on the product of coupling constants included in the vertices
of triangle loops and, in general, weakly depends on the parameters
of the $X(3872)$ resonance. Its value is a direct consequence of the
kinematics of the loops determined by the masses of the internal
particles. The isotopic symmetry prediction for $\mathcal{R}$ is
noticeably broken. The value obtained for $\mathcal{R}$ is a
specific prediction of the considered model, which gives an
opportunity to verify it experimentally.


\vspace*{0.3cm}
\begin{center} {\bf ACKNOWLEDGMENTS} \end{center}

We thank Professor. Dr. Zhiqing Liu for his interest in our
calculations of the $X(3872)\to\pi^+\pi^-\chi_{c1}$ decay rate. The
work was carried out within the framework of the state contract of
the Sobolev Institute of Mathematics, Project No. FWNF-2022-0021.




\begin{thebibliography}{99}
\bibitem{FN1} The reaction amplitude $D\bar D\to\pi\chi_{
c1}$ can be represented as $\varepsilon^{\mu*}[(q_{D}+q_{\bar
D})_\mu a+ (q_{D}-q_{\bar D})_\mu b]$, where $q_{D}$ and $q_{\bar
D}$ are the four-momenta of the $D$ and $\bar D$ mesons,
respectively, and $a$ and $b$ are the invariant amplitudes. We write
the polarization four-vector of the $\chi_{c1}$ as $\varepsilon^\mu=
\left( \frac{(\vec{p}_{\chi_{c1}},\vec{\xi})}{m_{\chi_{ c1}}},
\vec{\xi}+\frac{ \vec{p}_{\chi_{c1}}(\vec{p}_{\chi_{c1}},
\vec{\xi})}{m_{\chi_{c1}}(m_{ \chi_{c1}}+p^0_{\chi_{c1}})}\right)$.
Let us neglect here the mass difference of $D^0$ and $D^+$ mesons,
calculate the helicity amplitudes in the reaction center-of-mass
system, and expand them by the partial waves. Taking into account
Eq. (\ref{Eq6}) we obtain that near the $D\bar D$ threshold
$a\sim1/\mbox{GeV}^2$ and $b\sim(|\vec{q}||\vec{p}_{\chi_{c1}}|\cos
\theta)/\mbox{GeV}^4$, where $\theta$ is the angle between the
vectors $\vec{q}$ and $\vec{p}_{\chi_{c1}}$. This implies Eq.
(\ref{Eq7}).
\end{thebibliography}
\end{document}